\documentclass[preprint]{elsarticle}
\makeatletter
\def\ps@pprintTitle{%
 \let\@oddhead\@empty
 \let\@evenhead\@empty
 \def\@oddfoot{\centerline{\thepage}}%
 \let\@evenfoot\@oddfoot}
\makeatother
\usepackage{graphicx}
\usepackage{latexsym}
\usepackage{amsfonts}
\usepackage{caption}
\usepackage{color}
\usepackage{float}
\usepackage{amsmath,bbm}
\usepackage{algpseudocode}
\usepackage[utf8]{inputenc}
\usepackage[linesnumbered,algoruled,boxed,lined]{algorithm2e}
\usepackage{algorithmicx}
\usepackage{booktabs}
\usepackage{hyperref}
\SetKwRepeat{Do}{do}{while}%
\usepackage{amsthm}

\theoremstyle{definition}

\begin{document}

\begin{center}
The Version of Record of this manuscript \\
has been published and is available in \\
\emph{Complexity (2020)} \\
\url{https://doi.org/10.1155/2020/1047369}
\end{center}

\begin{frontmatter}
\title{Decentralizing Coordination in Open Vehicle Fleets for Scalable and  Dynamic Task Allocation}
 \author[Douai]{Marin~Lujak}
\ead{marin.lujak@imt-lille-douai.fr}
\author[Stefano]{Stefano~Giordani}
\ead{stefano.giordani@uniroma2.it}
\author[Andrea]{Andrea~Omicini}
\ead{andrea.omicini@unibo.it}
\author[Sascha]{Sascha~Ossowski}
\ead{sascha.ossowski@urjc.es}
\address[Douai]{CERI Numeric Systems, IMT Lille Douai,  59508 Douai,  France}
\address[Stefano]{Dip. Ingegneria dell'Impresa, University of Rome ``Tor Vergata'',  00133 Rome, Italy}

\address[Andrea]{Department of Computer Science and Engineering (DISI) Alma Mater Studiorum — Universit\`{a} di Bologna,  47522 Cesena, Italy}
\address[Sascha]{Centre for Intelligent Information Technologies (CETINIA), University Rey Juan Carlos, 28933 Madrid, Spain}

\begin{abstract} 
One of the major challenges in the coordination of large and open collaborative and commercial vehicle fleets is dynamic task allocation. Self-concerned individually rational vehicle drivers have both local and global objectives, which requires coordination using some fair and efficient task allocation method. In this paper, we review the literature on scalable and dynamic task allocation   focusing on deterministic and dynamic two-dimensional linear assignment problems.
We  focus on multi-agent system representation of open vehicle fleets where dynamically appearing vehicles are represented by  software   agents that should be allocated to a set of dynamically appearing tasks.  We give a comparison and critical analysis of recent research results focusing on centralized, distributed, and decentralized solution approaches. Moreover, we propose mathematical models for dynamic versions of the following assignment problems well-known in combinatorial optimization:
the assignment problem, bottleneck assignment problem, fair matching problem, dynamic minimum deviation assignment problem,  $\sum_{k}$- assignment problem, the semi-assignment problem, the assignment problem with side constraints,  and the assignment problem while recognizing agent   qualification; all while  considering the main aspect of open vehicle fleets: random arrival of tasks and vehicles (agents) that may become available after assisting previous tasks or by participating in the fleet at times based on individual interest.
\end{abstract}
\end{frontmatter}

\begingroup

\section{Introduction}\label{Introduction}
Open collaborative vehicle fleets composed of autonomous self-interested system participants are ever more widespread. However, even though the drivers are autonomous and self-interested, the authority and the ownership of these systems today remains centralized in terms of management, control, and access. 
The trend seems to be an ever increasing access to mobility and last-mile services for the average person at the cost of relying on just a few (centralized) worldwide enterprises.
State-of-the-art algorithms for the allocation of tasks to vehicle fleets solve customer requests in very large fleets in almost near real-time, but they seem to be limited to centralized systems.
Centralization here can be a source of: failure  (a single bottleneck of the system), obsolete information due to significant computation delay while processing ever increasing quantity of data, privacy evasion, and mistrust if the interests of the enterprise mismatch  the users' interest.

Distributed Decision-Making (DDM)  obviously resolves the drawbacks of centralized systems.
The multitude of the connected smart devices of the vehicles' drivers and customers makes it possible to combine their potential and to coordinate fleets at a scale exceeding spatial and computational boundaries. This potential can be exploited for the benefit of the fleet system as a whole as well as for the interest of individual vehicle drivers and customers.

The decision-making authority in the DDM is distributed throughout a system, and the decisions are taken locally based on the local and shared global information and the interactions of an individual with the rest of the system and with the environment. Here, each fleet participant is modelled as an autonomous collaborative individually-rational software agent installed on a user's smart device. The agent has only a local vision of the fleet and it needs to cooperate with other agents in order to find the allocation of dynamically appearing tasks faced by the whole fleet.

The behaviour of the fleet as a whole is a result of inter-vehicle coordination. Distributed task allocation strongly contributes to the shift of knowledge and power from the individual (fleet owner) to the collective (vehicles composing the fleet). A desired behaviour of the fleet emerges from the identifiable interest of its participating vehicles, their beliefs, and collective actions and, as such, is a shift away from the hierarchical organizational paradigm (see, e.g., \cite{horling2004survey}). A major challenge  is the identification of a right decision maker for each part of the problem, timely exchange of relevant and up-to-date information among vehicle agents, and modelling of complex relations in such a multi-agent system. 
A trade-off between the amount of computation and the quality of the solution is often necessary. Moreover, minimizing the overhead of communication required to converge to a desirable global solution is desirable.

\emph{Decentralized} coordination algorithms may be the means to obtain scalability for task allocation in the context of large-scale open fleets. Here, each self-concerned (vehicle, driver or courier) agent aims at achieving a desired local objective based on a limited local information and by communicating with the rest of the fleet and interacting with the environment.
Due to the limited local information, one of the drawbacks of decentralization is lack of control of the emerging fleet behaviour that cannot be predicted with certainty.
Moreover, to facilitate cooperation, assuming individually rational agents, we have to consider efficiency and fairness.
How to balance decentralization and centralization to improve system performance is much investigated but still not a completely-solved question.

\paragraph{Contribution} In this work, we present a survey on Multi-Agent System (MAS) coordination mechanisms for computationally complex dynamic  (one-on-one) task allocation problem (DTAP) and its variations for open vehicle fleet applications.
These problems may be modelled by a variety of deterministic and dynamic two-dimensional linear assignment problems, i.e., the problems regarding the assignment of two sets that may be referred to as ``agents'' and ``tasks'' with at most one task per agent and one agent per task, where the tasks appear dynamically and the task assignment is fully determined by the (cost, profit or revenue) parameter values and the initial conditions. We extend  mathematical models of the variations of the static task assignment problem  to their dynamic counterparts in open vehicle fleet scenarios considering, among others, self-interested and individually rational vehicle drivers, time restrictions, fairness, agent qualification and personal rank.

We identify some of the main scalable solution methods, i.e., coordination mechanisms, that can be put at work to solve these  problems. We investigate the theoretical scalability of these approaches and introduce a taxonomy to classify them in terms of the   level of inter-dependence in decision-making available to individual vehicles and customers during the coordination process (centralized, distributed, decentralized coordination). Our intention here is not to perform exhaustive search nor to identify the most scalable solution procedure. Contrarily, we  identify and mathematically model the  variations of the dynamic task assignment problem applicable to the studied fleet task allocation  contexts  and provide general scalability characteristics of their solution approaches. Our intention is to make it easier for a researcher to solve some variation of the task allocation problem in  large-scale open vehicle fleets by describing state-of-the-art solutions and their theoretical scalability results.

Even though some works exist that include reviews of the state of the art in multi-agent task allocation (see, e.g., \cite{chevaleyre2006issues,ahuja2017task,tang2010survey,jiang2012rich,jiang2019group}) and in vehicle fleet coordination (see, e.g., \cite{mariani2020coordination,billhardt2014dynamic,bielli2011trends}) or ridesharing optimization (see, e.g., \cite{agatz2012optimization,furuhata2013ridesharing}), none of them addresses one-on-one dynamic task assignment problems in open vehicle fleets.  In addition, a few approaches apply methods of multi-agent task allocation to the field of vehicle fleet coordination (see, e.g., \cite{billhardt2015towards}) but, to the best of our knowledge, there is no systematic survey combining both fields. 

The paper is organized as follows. In Section \ref{motivation}, we discuss some relevant concepts in the context of coordination for dynamic task allocation in open systems with the focus on distribution and decentralization of decision-making.
In Section \ref{taskAssignment}, we present mathematical models of various static and dynamic task assignment problems applicable in the open vehicle fleet context.
Centralized, distributed, and decentralized state-of-the-art  solution  methods and mechanisms  for the problems presented in Section \ref{taskAssignment} are discussed in Section \ref{scalabilityTheory}.
We conclude the paper emphasizing open issues and challenges for possible future research directions  in Section \ref{Challenges}. 
\section{Coordination in Open Vehicle Fleets}\label{motivation}
In this Section, we introduce some key concepts and characteristics of the target domains related to decentralizing coordination for scalable and dynamic task allocation.  The coordination problem arises   due to the distributed nature of the control exercised by the fleet's vehicles.

Generally, coordination may be defined as ``the process of organizing people or groups so that they work together properly and well''\footnote{\url{https://www.merriam-webster.com/dictionary/coordination}}.
By the coordination in   open vehicle fleets for task allocation, we refer to the organization and management of decision-making within the fleet with the aim to improve given key performance indicator(s) of a fleet's task allocation.

The topics of coordination and task allocation are the object of studies in multiple disciplines---e.g.,  operations research, economics, and computer science.
The corresponding definitions and related concepts may vary based on the specific discipline at hand.
In the so-called field of coordination models and languages, for instance, the focus is on the general-purpose abstractions (so-called \emph{coordination media}) that can be generally used to model and engineer the patterns of interaction between computational agents---with no specific reference to a particular application scenario or coordination problem.
In our survey, and in the following, we focus on the specific issues of dynamic task allocation and   distributed/decentralized coordination, with a particular emphasis on open vehicle fleets.

\subsection{Fleet coordination}
We consider  the context with cooperative vehicles in a large vehicle fleet, which functions as an organization that constrains the cooperation schemes within it.
The coordination problem here can be tackled from a bottom-up point of view, considering the emergence of global properties from the  inter-fleet direct vehicle to vehicle communication and fleet-environment interaction.

For simplicity and without loss of generality, we consider a two-dimensional space in which tasks may appear randomly at any location in space and time while the vehicles circulate through a transportation network within the space to reach them.
 Each vehicle can have three states: \emph{idle}, in which a vehicle is waiting for the assignment of a task, \emph{assigned} in which a vehicle is assigned to a task but has still not reached the task, and \emph{assisting} in which the vehicle has reached its assigned task and is assisting it. Only idle and assigned vehicles can be assigned or reassigned from one task to another. Once assigned, the vehicles start moving towards their assigned task. A task is considered completed once when it is reached and assisted by a vehicle.

Given a dynamically changing set (fleet) of idle and assigned vehicles, a dynamically changing set of  randomly appearing tasks, and a cost function of the assignment of each task to every idle and assigned vehicle (e.g., the  distance or time traveled or a given execution cost), the objective is to dynamically assign  these vehicles to tasks in a given time horizon reaching a globally minimum cost assignment considering that each task must be performed by exactly one vehicle.

Coordinating the vehicles in this respect requires that they find the globally best allocation in a distributed or decentralized way and resolve conflicts that violate local constraints. An efficient strategy in this context is a  dynamic   (re-)assignment of the vehicles in the fleet to the tasks as they appear. The vehicles require continuous communication and processing for task allocation. The coordination system must ensure a balanced use of shared resources, such as, e.g., Vehicle to Cloud (V2C) communication bandwidth and vehicle processing capacities.

V2C communication is limited in bandwidth and latency; so is the vehicle processing capacity.  Coordination strategies that ignore these communication and computation constraints may fail to find a fleet's action plan in close to real time and thus may be inapt for the application in   real-time fleets (see, e.g., \cite{zhu2018fog}).
These fleets require both autonomous and collaborative behaviors since vehicles have localized viewpoints, knowledge, and control and lack the overview of the global data integrated from various locations beyond their local capabilities. Such a  dynamic context requires for coordination-fault detection that indicates if the coordination exists within the fleet (see, e.g., \cite{lindner2009representation}). Once a coordination fault is detected, a coordination recovery process can begin in which cooperation can be rebuilt.

Vehicle fleets that rely on one-on-one vehicle-task assignment are, for example, rescue fleets (see, e.g., \cite{pujol2015efficient}), ride hailing and taxi service (see, e.g., \cite{billhardt2017coordinating}), ambulance assistance of urgent out-of-hospital patients (see, e.g., \cite{lujak2016distributed}),  and home-delivered restaurant hot meal services (see, e.g., \cite{ulmer2017restaurant}).
Ride hailing  and restaurant hot-meal delivery services are examples of open vehicle fleets that use online on-demand service platforms (see, e.g., \cite{taylor2018demand}) to allocate  in real-time customers and independent private  vehicle owners, drivers or couriers,  using their personal vehicles.
These platforms usually exploit sensor and GPS data to track the delivery process in real time \cite{dai2018information}.

Our focus is on the dynamic scenario with non-recurring  prearranged and  spontaneously requested single rider (customer), single driver  trips with at most one pickup and delivery  for each rider and driver.
Dynamically appearing riders (customers)  should be allocated to drivers in a one-on-one manner.
Before the allocation, in ride hailing, a customer chooses the driver based on the time of arrival and the price of the ride. In case of hot meal delivery, the system gives an estimated delivery time to the customer and assigns a courier that meets such an estimate.

\emph{Coordination} here is the key issue, including the stages of communication, resource allocation, and agreement.   The  allocation of the dynamically appearing customers over time needs to be performed in real-time and it fails if not completed within a specified deadline relative to an arrival of a customer; deadlines must always be met, regardless of the system load. Conventionally, the matching is based just on the rider's personal preferences and the nearby drivers' availabilities.
Reallocation of already-matched drivers to riders that are awaiting the service is not possible even if a more efficient matching exists.
At the end of each trip, every driver is available for a new rider allocation.

Speedy meal delivery services are constrained in geographic availability and timing. Usually, restaurants, riders, and  customers have access to the system through  an app. A customer detects his/her location and displays  restaurants that participate in the platform in the region of interest  and are open at the time.
Couriers participate in this open fleet context by delivering whenever they choose  and they may get paid on the individual delivery basis.
Once a customer requests a meal from a restaurant via his/her app, the corresponding delivery is assigned to a courier available nearby.
The courier picks up the delivery from the restaurant and delivers it to the customer.
After the delivery, a courier is available for new deliveries.

The allocation  of a courier to the customer is conventionally done based on the shortest arrival time to the restaurant (First-Come-First-Served strategy) and the availability of the courier; reallocation is not possible once the courier is allocated.
The  challenge here is to assign couriers to dynamically-appearing pickups and deliveries in order to maximize customer satisfaction (which can be measured in different ways, as explored in \cite{dai2018information}) without violating delivery times agreed at the time of the customer's hot meal request.

Task allocation problem in open vehicle fleets  considers both providers of transportation services (vehicle drivers) and their customers  and thus both of them may be considered active participants in the transportation process.
In the ride hailing scenario,   drivers are usually modelled as   agents and riders as tasks, while  in the hot meal delivery scenario, couriers are agents while meal deliveries are tasks.

Even though the ownership of most of open fleet systems today is centralized, not only customers, but also drivers with vehicles may appear dynamically and spontaneously in time and space influenced by a variety of   factors unknown in advance  such that it is reasonable to assume that they appear randomly.
In this \textit{dynamic task allocation} context, available vehicles are assigned to pending customers as they appear. Each agent and task is assumed to be characterized by a set of attributes that influences   the cost or profit resulting from an agent-task allocation. In this way, the \textit{task allocation} problem that  assigns tasks to agents in time is simplified to \textit{task assignment} problem focusing on the one agent - one task allocation at the time  (see, e.g., \cite{lujak2016distributed,billhardt2014dynamic2}).
Optimized and dynamic task (re-)assignment may considerably improve the performance of the fleet  while considering individual fairness and efficiency (see, e.g., \cite{billhardt2014dynamic2}). If dynamic courier (rider) reallocation is allowed, a substantial increase in efficiency may be observed, as in the case of ambulance allocation to out-of-hospital patients (see, e.g., \cite{billhardt2014dynamic,billhardt2014dynamic2,lujak2013coordinating}).

\subsection{Coordination models for open vehicle fleets}
Based on the ownership of the fleet, its structure, and the level of decentralizing coordination that we want to achieve in the fleet task allocation, we can design:
\begin{itemize}
  \item \textbf{a centralized coordination model}, where the task allocation problem   is solved in a single block by only one decision-maker (e.g., a single enterprise) having total control over and complete information about the vehicle fleet;
  \item \textbf{a distributed coordination model}, where the global task allocation problem is decomposed such that each customer is represented by an autonomous decision maker (agent) that may solve its own subproblem only with its own local decision variables and parameters. The allocation  of a limited number of vehicles (global constraints) is done through the interaction between competing customer agents and a vehicle fleet owner (a single autonomous agent) having available all the   fleet information.
   Customer agents that compete for the resources  are not willing to disclose their complete information but will share a part of it if it facilitates achieving their local objectives.
      The vehicle fleet owner agent is responsible of achieving globally efficient resource allocation by interacting with  customer agents usually through an auction. The problem decomposition here is done  to gain computational efficiency since customer agents can compute their bids in parallel. However, the resource allocation decisions  are still made by a single decision maker (vehicle fleet owner) with the requirement on synchronous bidding of customer agents (see, e.g. \cite{zavlanos2008distributed,giordani2010distributed,giordani2013distributed});
  \item \textbf{a decentralized cooordination model}, which further decentralizes the distributed model by allowing for multiple resource owner (vehicle) agents, multiple competing customer agents requesting the transportation service, and  asynchrony in decision-making. Customer agents compete for fleet's vehicles held by multiple resource owners   while each customer and resource owner agent has access only to its local information  with  no global information available. Therefore, they must negotiate resource allocation by running localized algorithms while exchanging relevant (possibly obsolete) information. Localized algorithms make the achievement of a desired global objective easier through simple local interactions of agents with their environment and other agents, with no need for a central decision maker. The decisions specifying these interactions emerge from local information. Fairness in resource allocation here plays a major role.
    The same as in the distributed model, an  agent is not willing to disclose its complete information but will share a part of it if it facilitates achieving its local objective. Resource allocation here is achieved by the means of a decentralized protocol.
\end{itemize}

Generally speaking, coordination is distributed when complex behavior within a system does not emerge due to the control of the system owner, but through interactions and communication of individual agents operating on local information, while sharing globally relevant knowledge.
This form of control is typically known as \emph{distributed control}, that is, control where each agent is equally responsible for contributing to the global, complex behaviour by acting properly on local information.
Agents are implicitly aware of the interaction rules through mechanisms that are based on the agent's interaction with other agents and the environment.
The system behaviour is then an emergent property of distributed coordination mechanisms (algorithms) that act upon agents, rather than the result of a control mechanism of a centralised system owner.
In decentralized algorithms, no global clock is assumed, no agent has complete information about the systems’ state, every agent takes decisions based only on local information, and failure of one agent does not prevent the system to continue running. An example is BitCoin: Instead of one central server owned and operated by a single entity, Bitcoin’s ledger is distributed across the Globe making it impossible to shut down, break-in, or hack as there is no single central bottleneck of the system.

Let us notice the main difference between distributed and decentralized coordination models.
Distributed coordination relies on local and shared (global) parameters and variables.
Local parameters and variables are private, whereas shared and global parameters and variables need to be shared among two or more agents---even among all the agents of the system.
If we assume self-concerned agents, resource owner can manipulate these parameters and variables or deceive agents in communicating their values to influence the individual decision making of each one of them and thus obtain the behavior of the system the resource owner wants.
This can be prevented by ensuring individual agent access to non-obsolete and truthful information---using e.g.\ blockchain technology. Reaching a globally optimal solution with quality of solution guarantees is then possible, contrary to the decentralized coordination case.  In the latter case, due to  the lack of the global non-obsolete and truthful   information, quality of solution guarantees generally do not exist. In general, solution approaches for decentralized coordination concentrate on finding a feasible (admissible) solution without quality of solution guarantees. Contrary to the distributed case most often studied in the operations research field where the emphasis is on the method's optimality gap,  decentralized coordination methods are mostly approximate heuristics-based methods  without quality of solution guarantees but with proven completeness, soundness, and termination.
	
Open vehicle fleets are intrinsically distributed systems since they comprise a multitude of  geographically distributed and mutually communicating customers' and vehicle drivers'  apps.
Traditionally, distributed systems refer to systems consisting of sequential processes (each one with an independent thread of control, possibly located on geographically distributed processors) that coordinate their actions by exchanging messages to meet a common goal (see, e.g., \cite{ghosh2014distributed,tanenbaum2007distributed}).  The common goal in this context is an efficient and cost-effective transportation service of the vehicle fleet while considering individual rationality, preferences and constraints whether it is of drivers, riders, or hot meal delivery customers. Quality of solution guarantees play a crucial role of sustainable competitive advantage in any transportation network company.

Distributed open vehicle fleets exhibit some clear strong points over their centralized counterparts.
First of all, they are more robust than their centralized counterparts  because they can rely on their intrinsic built-in redundancy. They can operate at a larger scale and assist more customers at once since they are aggregating vehicle capacity and customer throughput across all their individual vehicle drivers.
However, distributed open vehicle fleets also have to deal with inter-vehicle communication and coordination overhead that can sometimes make them slower or more difficult to control than their centralized counterparts.
Applying trustless distributed systems that are meant to operate in an adversarial environment, such as Bitcoin, in open fleets entails an additional overhead.

\section{Task assignment models for open vehicle fleets}\label{taskAssignment}
Assignment problems (APs) are among the earliest optimization problems studied in the operations research field. They involve optimally matching the elements of two or more sets, where the
dimension of the problem refers to the number of sets to be matched \cite{pentico2007assignment}.
For example, in two-dimensional assignment problems,  given is a set of agents $A$ and a set of tasks $T$ and we have to match (assign) tasks to agents. Tasks are assumed atomic, i.e., each task cannot be decomposed into subtasks and it can be completed by a single vehicle.
In general, two-dimensional assignment problems can be solved in polynomial time, while  $d$-dimensional assignment problems, with $d > 2$, in general are NP-hard  (see, e.g., \cite{burkard1999linear}).

We distinguish between the static and dynamic assignment problems (see, e.g., \cite{spivey2004dynamic}). The former refer to the assignment of a  set of tasks to a set of agents in a given static environment in which the problem data does not change  during the planning horizon, while in the dynamic task assignment problems, both agents and tasks may appear and disappear dynamically over time.
In the open vehicle fleet setting, agents  can be in one of the following three states: \textit{idle}, \textit{assigned} without still having reached the customer, or \textit{assisting a customer}, and only idle and assigned agents that have still not reached their customers can be (re)assigned to unassisted tasks. In general,  agents are assumed renewable, i.e., after completing a task, an agent's state changes from \textit{assisting a customer} to  \textit{idle} and it becomes assignable again to customers (tasks) that have not been assisted yet. This  is a special case of a more general computationally complex dynamic vehicle routing problem (DVRP) in which, for each (vehicle) agent, we find  a minimum cost route that visits a  dynamically changing set of  tasks (customers) \cite{pillac2013review}.
Due to the high computational complexity, myopic algorithms are the most usual solution approaches for DVRP. For simplicity, we can assume that agents are nonrenewable, i.e., an agent can be assigned only to one task; if, after completing a task, it is still available for new task assignment, it appears as a new agent.

The static and deterministic AP is a computationally easy problem, which allows us (in theory) to find an optimal solution in close-to real-time  (in the nonrenewable agent case).
Dynamic AP can be solved by (suboptimal) myopic approaches that consider only the information available at the present time with no consideration for future events and possibly reassign tasks among idle and already assigned agents to improve the system's efficiency (see, e.g., \cite{billhardt2014dynamic,lujak2016distributed,billhardt2014dynamic2,lujak2013coordinating}).
However, in the case where tasks are not randomly appearing, this approach can be significantly improved by considering future developments.

\subsection{Static Task Assignment}
Based on the categorization of the  AP models presented in \cite{pentico2007assignment}, in this section, we consider  the classic assignment problem and its variations relevant in the open fleet vehicle task assignment considering self-interested and individually rational vehicle users whose tasks can be performed simultaneously: the classic linear assignment problem (LAP), assignment problem recognizing agent qualification (APRAQ), the bottleneck assignment problem (BAP), the fair matching problem (FMP), the minimum deviation assignment problem (MDAP), the $\sum_k$-assignment problem ($\sum_k$-AP), the semi-assignment problem (SAP), and the assignment problem with side constraints (APSC). In Figure \ref{Framework}, we give a framework for easier understanding of the characteristics of both the static and dynamic version of these problems.

For self-completeness of this article,  we bring in the following the descriptions of these problems. Considering that the number of publications concerning assignment problems is enormous, the
references in this section constitute only a very limited part of them.
For the details and other assignment problem variations the reader is referred to \cite{pentico2007assignment}.

\begin{figure}[ht]
\begin{center}
\includegraphics[width=0.8\columnwidth]{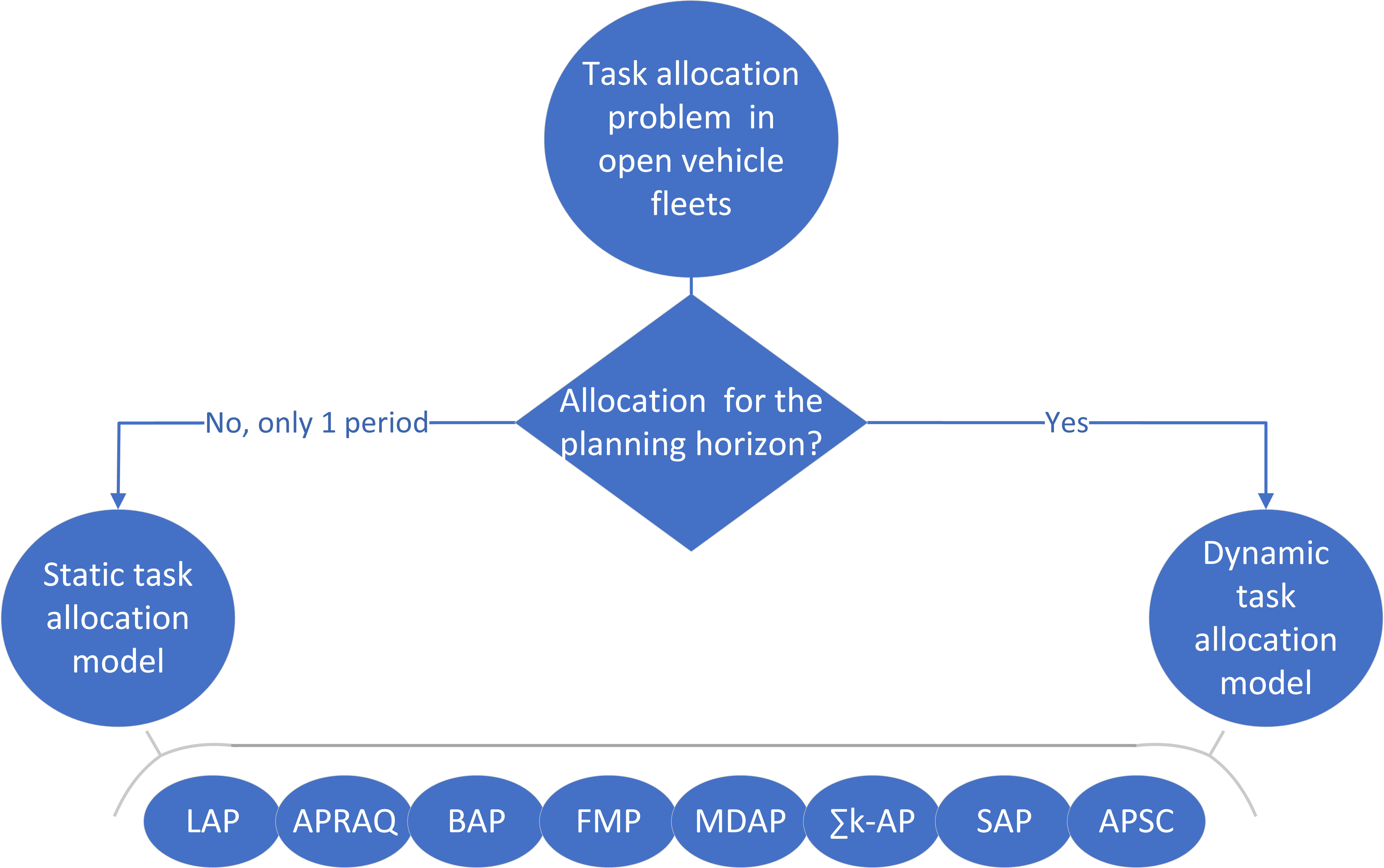}
\caption{Static and dynamic task assignment problems in open vehicle fleets}
\label{Framework}
\end{center}
\end{figure}

\paragraph{Classic (linear) assignment problem (LAP)}
The static classic linear assignment problem involves two sets of the same size and consists of finding, in a weighted complete bipartite graph, a perfect matching in which the sum of weights of the matched edges is as low as possible, i.e., a \textit{minimum-weight perfect matching}. Perfect weighted matching implies that each node must be matched to some other node by minimizing the total cost of the arcs in the (perfect) matching.

The classic linear assignment problem (LAP) can be defined  as follows: Given a weighted complete bipartite graph $G = (A \cup T, E)$ with two vertex sets $A$ and $T$, with $n = |A| = |T|$, and an edge set $E = A \times T$, with edge weights $c_{ij}$ on edge $(i,j) \in E$,
find a minimum weight perfect matching of $G$, i.e., a perfect matching among vertices in $A$ and vertices in $T$ such that the sum of the costs of the matched edges is minimum. An edge $(i,j) \in E$ is matched if  two extreme vertices $i$ and $j$ are mutually matched, and a matching is perfect if every vertex $i$ of $A$ is matched (assigned) exactly to one vertex $j$ of $T$, and viceversa.
The LAP is equivalent to the weighted bipartite matching, since we may assume that the bipartite graph is always complete by letting the weights of the edges that are missing being sufficiently large.
If $|A| \neq |T|$,  we can add a number of dummy nodes to the set with lower cardinality and connect them by dummy arcs of zero cost to the other set. The number of dummy nodes should be  sufficient to balance the cardinalities of the two sets.

The LAP is equivalent to the maximum weighted bipartite matching (with edge weights $w_{ij} \geq 0$), since we may assume that the bipartite graph is always complete by letting the weights of the edges that are missing being sufficiently large. Furthermore, also in this case we can assume that the two vertex sets of the bipartite graph have the same size. At this point we can reformulate the problem as a minimization problem by considering costs $c_{ij} = W - w_{ij}$, where $W$ is larger than the maximum of the $w_{ij}$, and hence this problem corresponds to the LAP.

The LAP is a special case of the transportation problem assuming an equal  number of supplier agents and customer agents and each one with their unitary  supply and unitary demand, respectively. The transportation problem is one of special cases of the minimum cost flow problem together with, e.g., the shortest path problem and the max flow problem. While it is possible to solve this problem using the Simplex algorithm, specialized  algorithms   take advantage of its special network structure and are thus more efficient.

From the multi-agent systems' point of view, in the assignment problem, a number of agents need to be assigned to a number of tasks based on the given cost of agent-task assignment. In general, each agent can be assigned to any task. In  case an agent is not capable of performing a task, a given agent-task assignment cost is modelled as a very large number.  All tasks should be performed with the objective to minimize the   total cost of the assignment such that exactly one agent is assigned to each task and  exactly one task  to each agent.
The mathematical formulation of the problem is:
\begin{equation}\label{APobjective}
\min \sum_{i,j}c_{ij}x_{ij}
\end{equation}
subject to
\begin{equation}\label{tasks}
\sum_{i=1}^n x_{ij}=1,\;\;\; \forall\; j \in T,
\end{equation}
and
\begin{equation}\label{agents}
\sum_{j=1}^n x_{ij}=1, \; \; \; \forall\; i \in A,
\end{equation}
\begin{equation}\label{non-negat}
    x_{ij} \in \{0,1\}, \; \; \; \; \forall\; i \in A,\;j \in T.
\end{equation}

Constraints (\ref{tasks}) ensure that every
task is assigned to only one agent and constraints (\ref{agents}) ensure that every agent is assigned
to only one task.

The structure of the problem, i.e., the total unimodularity of the constraint matrix, makes the binary requirements on the variables unnecessary. In fact, in this case, it can be proven that the linear relaxation has always an optimal binary solution (see, e.g., \cite{lawler2001combinatorial,papadimitriou1982combinatorial}) and, therefore, the LAP is a linear programming (LP) problem.

\paragraph{The classic assignment problem recognizing agent qualification (APRAQ)}

Caron et al. in \cite{caron1999assignment} propose a mathematical model in which not every agent is qualified
to do every task, and the objective is utility maximization:
\begin{equation}
\max \sum_{i,j}p_{ij}x_{ij}
\end{equation}
subject to
\begin{equation}\label{qualifiedTasks}
\sum_{i\in A} q_{ij} x_{ij} \leq 1,\;\;\; \forall\;j \in T\;,
\end{equation}
and
\begin{equation}\label{qualifiedAgents}
\sum_{j\in T} q_{ij} x_{ij}\leq 1, \; \; \; \forall\; i \in A\; ,
\end{equation}
\begin{equation}
    x_{ij}\in \{ 0,1\}\;, \forall i \in A, \;j \in T,
\end{equation}
where parameter $q_{ij}$ = 1 if agent $i$ is qualified to perform task
$j$, 0 otherwise, parameter $p_{ij}$ is the utility of assigning agent
$i$ to task $j$ (with $p_{ij}$ = 0 if $q_{ij}$ = 0), and variable
$x_{ij}$ = 1 if agent $i$ is assigned to task $j$, 0 otherwise.
Constraints (\ref{qualifiedTasks}) ensure that no more than one qualified agent is assigned to any task, while  constraints (\ref{qualifiedAgents}) guarantee that each agent is assigned
to not more than one task.

The classic assignment problem does not consider fairness.  The solution of classic AP (\ref{APobjective})-(\ref{non-negat})  maximizes utilitarian social welfare (see, e.g., \cite{moulin2004fair}), but it may be unfair and unsatisfactory since there may be one or more agents with a much higher  task cost than the rest. This is why it is best applied to centralized open vehicle fleets with a single owner of the fleet's vehicles that is interested in the minimization of the overall cost of the fleet's operation costs but not in how they are distributed among the vehicles.

\paragraph{Bottleneck assignment problem (BAP)}

To resolve the issues with fairness and workload distribution, we may  minimize maximum cost among the individual agent-task assignments and thus maximize the system's egalitarian social welfare (see, e.g., \cite{burkard2009}). The mathematical program for the BAP is as follows:
Minimize $\max_{i,j}\{c_{ij}x_{ij}\}$ or minimize $\max_{i,j} \{c_{ij}| x_{ij}=1\}$
subject to constraints (\ref{tasks})--(\ref{non-negat}) and definitions of
 the LAP.

Note that here the integrality requirements cannot be relaxed.
Contrary to the classic AP model, the BAP  model pursues the objective of fairness among agents.
It is based on the optimization of the worst-off performance and provides a good solution when the minimum requirements of all agents should be  satisfied. However, only the most costly agent-task assignment influences the objective function, while the contribution of  the rest  of the agents is ignored. For this reason, this approach deteriorates the system efficiency and thus, the system's utilitarian social welfare.

\paragraph{The  fair matching problem (FMP)}

The fair matching problem  minimizes the difference between the maximum and minimum assignment values \cite{martello1984balanced}:

Minimize $\max_{i,j}\{c_{ij}|x_{ij}=1\}-\min_{i,j}\{c_{ij}|x_{ij}=1\}$
subject to the same constraints and definitions as in the classic AP.

This formulation of fairness is not unique. Sun and Yang (2003) in \cite{sun2003general} study the concept of fair and optimal allocations.
They define an allocation to be  fair and optimal if it is envy-free and the sum of compensations is maximized, subject to the compensation assigned to each object is less than or equal to the maximum compensation limit. They prove that fair and optimal allocations exist and demonstrate that the fair and optimal allocation mechanism achieves efficiency, fairness and strategy-proofness simultaneously. \cite{andersson2009general} demonstrates that it is also coalitionally strategy-proof, i.e., it is not possible for any agent or any coalition of agents to successfully manipulate the allocation rule.

\paragraph{The minimum deviation assignment problem (MDAP)}

The objective here is to minimize the difference between the maximum and average assignment costs:\\

\begin{equation}
Minimize \;\;\min\{n,m\}\times \max_{p,q}\{c_{pq}x_{pq}\}-\sum_{i=1}^n \sum_{j=1}^m c_{ij}x_{ij}
\end{equation}
or  to minimize the difference between the average  and minimum assignment profit:
\begin{equation}
Minimize \;\;\sum_{i=1}^n \sum_{j=1}^m p_{ij}x_{ij}-\min\{n,m\}\times \min_{s,t}\{p_{st}x_{st}\},
\end{equation}
subject to constraints (\ref{tasks})--(\ref{non-negat}). Here, $n$ is cardinality of agent set $A$, and $m$ of task set $T$, other  definitions are the same as in the LAP \cite{duin1991minimum,gupta1988minimum}.

\paragraph{The $\sum_k$-assignment problem ($\sum_k$-AP)}

Since there may be generally multiple different sets of assignments with the same minimum value for $\max\{c_{ij}x_{ij}\}$, the objective here is to find a set of assignments for
which the sum of the $k$ largest values is minimized. The BAP and LAP can be viewed as special cases of $\sum_k$-AP with $k=1$ and $k=n$, respectively.

A recent study on generic mixed integer problem with $\sum_k$ optimization is done by Filippi {\em et al.} \cite{filippi2019bridging}.

\paragraph{The semi-assignment problem (SAP)} This is the version of the assignment problem where every agent or task may not be unique. This results in a constraint matrix containing a number of rows or columns with equal  coefficients. Kennington and Wang in \cite{kennington1992shortest} show examples of such a problem in workforce  and project planning and scheduling as use case examples.
Here, constraints (\ref{tasks}) from the classic LAP are substituted by
\begin{equation}
\sum_{i=1}^m x_{ij}=d_j,\;\;\; \forall\;j,
\end{equation}
everything else  being the same as in the classic LAP for the situation in which there are $n$ agents and $m$ task categories. Here, $m \leq n$, and $d_j$ is the number of tasks in task group $j$ with $\sum_j d_j = n$.

Note that if also the agents are not unique and are clustered into agent groups, with $q_i$ agents in each group $i$, where $\sum_j d_j = \sum_i q_i$, the problem is equivalent to the transportation problem.

\paragraph{The assignment problem with side constraints (APSC)}
Classic assignment  problem can be solved by multiple centralized and efficient polynomial algorithms. However, by introducing side constraints, generally, this problem becomes NP-hard.
Side constraints may include budgetary limitations, degree of technical training of personnel, the rank of personnel, or time restrictions, that limit the assignment of agents to tasks.

Aggarval \cite{aggarwal1985lagrangean} introduces to the classical LAP problem an additional knapsack-type constraint
\begin{equation}\label{knapsackConstraint}
\sum_{i,j}r_{ij}x_{ij}\leq b,
\end{equation}
where  $r_{ij}$ is the amount of resource   used if agent $i$ is assigned to task $j$ and
$b$ is the amount of a resource   available. Adding constraint (\ref{knapsackConstraint}) to LAP results in a Resource Constrained Assignment Problem (RCAP), which is a knapsack problem under perfect matching over a bipartite network. Constraint
(\ref{knapsackConstraint}) deranges the unimodularity of the LAP set of constraints so that the optimal solution of the linear relaxation of the problem is no more always within the values $\{0,1\}$ and, hence, integrality constraints cannot be relaxed.
The resulting problem  belongs to the class of NP-complete problems for which no polynomially-bounded algorithm is likely to exist (see, e.g., \cite{aggarwal1985lagrangean}).

Mazzola and Neebe
\cite{mazzola1986resource} present a general model for the assignment problem with side constraints that generalizes the General Assignment Problem (GAP) (see, e.g., \cite{cattrysse1992survey}) and adds the following  constraints to either the classic LAP model or the classic LAP recognizing agent qualifications:
\begin{equation}\label{sideConstraints}
\sum_{i,j}r_{ijk}x_{ij}\leq b_k, \;\; \forall k,
\end{equation}

\noindent where $r_{ijk}$ is the amount of resource $k$ used if agent $i$ is assigned to task $j$ and $b_k$ is the amount of resource $k$ available.

By side constraints, we can model   drivers that belong to different seniority classes and customers that have different  priority levels. Seniority constraints impose for the solution to be such that no unassigned agent can be assigned to a task unless an assigned agent with the same or higher seniority becomes unassigned, while priority constraints specify that the solution must be such that no unassigned task can become assigned without a task with the same or higher priority becoming unassigned \cite{caron1999assignment}.

\subsection{Dynamic task assignment}
In this section, we propose extensions of the static assignment problem models presented previously   to the dynamic  versions in which new agents and tasks may enter the system in each time period and the costs or profits of agent-task assignment are updated in (close-to) real-time.
This problem is similar to the on-line bipartite matching problem, in which  tasks that appear in sequence should be assigned to the agents immediately as they appear.
Relating to the previously presented terminology of the static AP, a set of available (idle and assigned)  agents $A$ (that are not assisting any customer) is  known in the given weighted bipartite graph $G = (A \cup T, E)$. Tasks in $T$ (along with their incident edges) arrive online. Upon the arrival of a task $j \in T$, we must assign it  to one of agents $i \in A$ with an existing edge $(i,j) \in E$. At all times, the set of matched edges must form a (feasible) matching, i.e.,  each agent should be matched with at most one task and viceversa. In case of different cardinalities of the two sets, to balance the two, dummy elements are added to the set with lower cardinality.

We assume random arrivals of customer demands (tasks) over time. In open fleets, we also assume that agents (drivers and couriers)  either  become available randomly after assisting previous tasks (customers) or by entering and leaving the fleet based on personal interest, available time, and/or other individual constraints and preferences. Given are attribute parameters both  for agents and tasks that define their main characteristics in terms of the assignment.

We consider deterministic on-demand task allocation where the (re-)assignment of vehicles (agents) to  tasks is performed as soon as  a new vehicle or task enters the system. Close to real time reassignment is beneficial here since the parameters and variables of the assignment problem are perfectly known.

Spivez and Powell \cite{spivey2004dynamic} propose a Markov decision process model for the dynamic assignment problem.  In this paper, inspired by their work,  we propose mathematical programming models for the variations of the static task assignment described in the previous section while respecting agent-task taxonomy used previously in this paper.

The decisional variables in the dynamic AP receive a third index such that:
\begin{equation}
  x_{i j \tau}=
\begin{cases}
  1, & \text{if task } j \in T \text{ is assigned to agent } i \in A \text{  at period } \tau \in \mathcal{T}\\
  0, & \text{ otherwise.}
\end{cases}
\end{equation}

Moreover, we introduce two additional binary variables $\alpha_{\tau i}$ and $\beta_{\tau j}$, for all $i \in A$, $j \in T$ defined as follows.
\begin{equation}
  \alpha_{i\tau }=
\begin{cases}
1, & \text{if agent } i \in A \text{ is known and available for assignment in period } \tau \\
  0, & \text{otherwise.}
\end{cases}
\end{equation}

\begin{equation}
  \beta_{j\tau }=
\begin{cases}
  1, & \text{if task } j \in T \text{ is known and available for assignment in period } \tau \\
  0, & \text{otherwise.}
\end{cases}
\end{equation}

Let $\mathcal{T}$ be a set of consecutive time periods of the planning time horizon.
The mathematical formulation of the deterministic and dynamic  LAP problem  considering utility maximization is then given by:
\begin{equation}\label{DAP}
Z=\max \sum_{\tau \in \mathcal{T}}\sum_{i \in A}\sum_{j \in T} p_{i j \tau} x_{i j \tau}
\end{equation}

subject to:

\begin{equation}\label{agentsLimit}
\sum_{j \in T} x_{i j \tau} \leq \alpha_{i\tau }, \; \forall i, \tau
\end{equation}

\begin{equation}\label{tasksLimit}
\sum_{i \in A} x_{i j \tau} \leq \beta_{ j\tau } \;\forall j, \tau
\end{equation}

\begin{equation}\label{conservationOfAgents}
\alpha_{i,\tau+1} = \alpha_{i\tau }-\sum_{j \in T} x_{i j \tau}+\hat{A}_{ i,\tau+1}, \;\forall  i, \forall \tau \in \{1, \ldots, |\mathcal{T}|-1\}
\end{equation}

\begin{equation}\label{conservationOfTasks}
\beta_{j,\tau+1} = \beta_{j\tau}-\sum_{i \in A} x_{i j \tau}+\hat{T}_{j,\tau+1}, \; \forall j, \forall \tau \in \{1, \ldots, |\mathcal{T}|-1\}
\end{equation}

\begin{equation}\label{initialCondAgents}
\alpha_{i,1} =  \hat{A}_{ i,1}, \;\forall  i
\end{equation}

\begin{equation}\label{initialCondTasks}
\beta_{j,1} =  \hat{T}_{j,1}, \; \forall j
\end{equation}

\begin{equation}\label{nonNeg1}
x_{i j \tau}\in \{0,1\}, \forall i \in A, j \in T, \tau \in \mathcal{T}
\end{equation}

\begin{equation}\label{nonNeg2}
\alpha_{i \tau} \in \{0,1\},  \forall i \in A, \tau \in \mathcal{T}
\end{equation}

\begin{equation}\label{nonNeg}
\beta_{j \tau}\in \{0,1\}, \forall j \in T, \tau \in \mathcal{T},
\end{equation}

where $p_{ij\tau}$ is the utility of assigning agent $i$ to task $j$ at period $\tau$ (note that it may vary through time) and  $\hat{A}$ and $\hat{T}$ are given parameters such that:

\begin{equation}
  \hat{A}_{i\tau}=
\begin{cases}
1, & \text{if agent } i \in A \text{ enters into set $A$ (the fleet) in period } \tau \\
  0, & \text{otherwise.}
\end{cases}
\end{equation}

\begin{equation}
  \hat{T}_{j\tau}=
\begin{cases}
  1, & \text{if task } j \in T \text{ becomes known in period } \tau \\
  0, & \text{otherwise.}
\end{cases}
\end{equation}

Moreover, based on the assumption of nonrenewable agents and tasks, we assume that:  $\sum_{\tau \in \mathcal{T}}\hat{A}_{i\tau} \leq 1$ and $\sum_{\tau \in \mathcal{T}}\hat{T}_{j\tau } \leq 1$, i.e., every agent and task are unique and enter into the fleet and thus become available for assignment only once.

The aim is maximizing the total utilitarian social welfare over the planning time horizon, which is achieved by maximizing the  assignment utility (\ref{DAP}) over all agent-task assignments in all periods of the planning time horizon. Constraints (\ref{agentsLimit}) guarantee that each available agent at time period $\tau$ is assigned to at most one task while unavailable agents cannot be assigned to any task.
Constraints (\ref{tasksLimit}) ensure that at most one agent is assigned to any available task while no agent can be assigned to any unavailable task.

Constraints (\ref{conservationOfAgents}) and (\ref{conservationOfTasks})
represent the dynamics of dependant variables $\alpha_{\tau i}$ and  $\beta_{\tau j}$, assuming that both agents and tasks disappear from the system at the end of the period when they are assigned.
Furthermore, constraints (\ref{initialCondAgents}) and (\ref{initialCondTasks}) represent initial conditions of the problem, while the variable ranges are given by (\ref{nonNeg1})-(\ref{nonNeg}).

We can also consider cost minimization problem where we substitute (\ref{DAP}) with the following objective function
\begin{equation}\label{DAPMin}
Z=\min \sum_{\tau \in \mathcal{T}}\sum_{i \in A}\sum_{j \in T} c_{i j \tau} x_{i j \tau}
\end{equation}
subject to:
\begin{equation}\label{assignAllTasksorAgents}
\sum_{i \in A}\sum_{j \in T} \sum_{\tau \in \mathcal{T} }  x_{i j \tau} = n
\end{equation}
and   (\ref{agentsLimit})--(\ref{nonNeg}). Constraint
(\ref{assignAllTasksorAgents}) guarantees the assignment of all the tasks and/or agents in the planning time horizon, depending on the relative size of these two sets.

\paragraph{The dynamic classic assignment problem recognizing agent qualification}
Here,  the objective function is again the utility maximization (\ref{DAP}),
while constraints (\ref{agentsLimit}) and (\ref{tasksLimit}) are substituted by the following ones, everything else remaining the same as in the dynamic LAP:
\begin{equation}\label{DynamicqualifiedTasks}
\sum_{j\in T} q_{ij\tau} x_{ij\tau} \leq \alpha_{\tau i},\;\;\; \forall\;i, \tau
\end{equation}
and
\begin{equation}\label{DynamicqualifiedAgents}
\sum_{i\in A} q_{ij\tau} x_{ij\tau}\leq \beta_{\tau j}, \; \; \; \forall\; j, \tau
\end{equation}

where parameter $q_{ij\tau}$ = 1 if agent $i$ is qualified to perform task
$j$ at period $\tau$, 0 otherwise, parameter $p_{ij\tau}$ is the utility of assigning agent
$i$ to task $j$ at period $\tau$ (with $p_{ij\tau}$ = 0 if $q_{ij\tau}$ = 0), and variable
$x_{ij\tau}$ = 1 if agent $i$ is assigned to task $j$ at period $\tau$, 0 otherwise.
Constraints (\ref{DynamicqualifiedTasks}) guarantee that no more than one qualified agent is assigned to any task, while  constraints (\ref{DynamicqualifiedAgents}) ensure that each agent is assigned
to not more than one task.
Instead of the profit  maximization, here, we can introduce cost minimization by  substituting (\ref{DAP}) with (\ref{DAPMin}) and introducing (\ref{assignAllTasksorAgents}) into the constraint set.

\paragraph{The dynamic bottleneck assignment problem (DBAP)}
The objective function of the DBAP problem can be formulated   as follows: at each period $\tau \in \mathcal{T}$,
maximize $Z= \min_{i,j}\{p_{ij\tau}x_{ij\tau}\}$ or maximize $Z= \min_{i,j} \{p_{ij\tau}| x_{ij\tau}=1\}$.
 This maxmin problem can be  expressed by maximizing an additional variable $L$ that is a lower bound for each of the individual values $\{p_{ij\tau}|x_{ij\tau}=1\}$ as follows: $\max L$ subject to constraints $L \leq \sum_{j \in T} p_{ij\tau} x_{ij\tau}$ for all $i \in A_{\tau}$, $\tau \in \mathcal{T}$, and   (\ref{agentsLimit})--(\ref{nonNeg}) and definitions of
 the dynamic LAP.

\paragraph{The dynamic fair matching problem (DFMP)}
Here, at each period $\tau \in \mathcal{T}$,we minimize the objective function $\max_{i,j}\{c_{ij\tau}|x_{ij\tau}=1\}-\min_{i,j}\{c_{ij\tau}|x_{ij\tau}=1\}$ and
subject to constraints (\ref{agentsLimit})--(\ref{nonNeg}).
Similarly, we can minimize the difference between the maximum and minimum profit obtained among agents, i.e.,
$minimize \;\;(\max_{i,j}\{p_{ij\tau}|x_{ij\tau}=1\}-\min_{i,j}\{p_{ij\tau}|x_{ij\tau}=1\}$ and
subject to constraints (\ref{agentsLimit})--(\ref{nonNeg}).

\paragraph{The dynamic minimum deviation assignment problem (DMDAP)} At each period $\tau \in \mathcal{T}$, the objective function is as follows:
\begin{equation}
Minimize \;\; \min\{n,m\}\times \max_{p,q}\{c_{pq\tau}x_{pq}\}-\sum_{i\in A} \sum_{j\in T} c_{ij\tau}x_{ij\tau}
\end{equation}
or:
\begin{equation}
Minimize \;\;\sum_{i=1}^n \sum_{j=1}^m p_{ij}x_{ij}-\min\{n,m\}\times \min_{s,t}\{p_{st}x_{st}\},
\end{equation}
subject to constraints (\ref{agentsLimit})--(\ref{nonNeg}) and definitions of
 the minimum deviation assignment problem.

\paragraph{The dynamic $\sum_k$-assignment problem (D$\sum_k$-AP)}
Given parameter $k$,  objective function (\ref{DAP}) is modified to:
\begin{equation}\label{DSUMAP}
Z=\max \sum_{\tau \in \mathcal{T}}\sum_{i=1}^k\sum_{j \in T} p_{i j \tau} x_{i j \tau}
\end{equation}
subject to constraints (\ref{agentsLimit})--(\ref{nonNeg}) and definitions of
 the dynamic LAP.

\paragraph{The semi-assignment problem}
Here, constraints (\ref{tasksLimit}) from the dynamic LAP are substituted by
\begin{equation}
\sum_{i=1}^m x_{ij\tau}=d_{j} \beta_{j \tau},\;\;\; \forall\;j,\tau
\end{equation}
everything else  being the same as in the dynamic LAP for the situation in which there are $n$ agents and $m$ task categories, where $m \leq n$.

\paragraph{The assignment problem with side constraints}

Side constraints (\ref{sideConstraints}) here include also the time index:
\begin{equation}\label{dynamicsideConstraints}
\sum_{i,j}r_{ijk\tau}x_{ij\tau}\leq b_{k\tau} \;\; \forall k,\tau,
\end{equation}
\noindent where $r_{ijk\tau}$ is the amount of resource $k$ used if agent
$i$ is assigned to task $j$ at period $\tau$ and $b_{k\tau}$ is the amount of resource
$k$ available at period $\tau \in \mathcal{T}$. Constraints (\ref{dynamicsideConstraints}) are simply added to the formulation of the dynamic LAP.

\subsection{Bottomline}
To sum up,  in  Table \ref{tab:table-name}, we give the overview of the characteristics of the treated (static and dynamic) task assignment problems   related with: i) the kind of the social welfare they optimize (utilitarian, egalitarian, elitist, or a difference between them), ii) whether agents are qualified to perform only certain tasks or not, iii), including fairness or not, iv) whether the agents are considered homogeneous or not, v) time restrictions, vi) personal ranking, and vii) technical training.
\begin{table}\centering
\begin{tabular}{l|ccccccc} 
\toprule
    & Soc. & Agent  & Fairness & Unique & Time & Pers. & Tech. \\
     Model & welfare & qualif.  &  & ag./tasks & restr. & rank & train. \\\midrule
    LAP  & util. & no  & no & yes  & no & no &  no  \\
    APRAQ & util. & yes  & no & yes  & no & no & no   \\
    BAP  & egal. & no  & no & yes  & no & no  & no     \\
    FMP  & el. - eg. &no   & no & yes   & no & no & no \\\midrule
    MDAP  & el.- ut. & no  & no & yes  & no & no & no    \\
     & ut.- el. & no  & no & yes & no & no & no    \\
    $\sum_k$-AP  & egal. & no  & yes & no  & no & no & no     \\
    SAP & util. & no  & no & yes  & no & no  & no       \\ 
    APSC & util. & no  & no & no  & yes & yes & yes  \\\bottomrule
\end{tabular}
\caption{\label{tab:table-name}Characteristics of the discussed task assignment models}
\end{table}

Note that once we introduce additional constraints to the classic assignment problem,  the resulting model is, generally, no more resolvable in polynomial time and is highly computationally expensive.
Additionally, \emph{we consider tasks and agents that may be  known both at some future time period   and at the first period of the planning time horizon}. Therefore, we can use this model to coordinate task allocation for  planned tasks and agents  that  schedule their appearance in advance for some future time period, but also for the tasks and agents that need to be allocated    on short-notice or immediately as they get known and enter the system. To this aim, we must use     highly computationally efficient close-to real-time  solution approaches and, generally, exact methods do not suffice for this purpose. Therefore, we are obliged to use heuristic-based approximations.

\section{Coordination approaches in task allocation to fleet's vehicles }\label{scalabilityTheory}
In this section, we recall the main (coordination) solution  methods for the task allocation problem in open vehicle fleets in general and the treated assignment problems in particular, categorizing them in  centralized, distributed, and decentralized (Figure \ref{SolutionApproaches}), with special attention to those with the best time complexity. Recall that the static classic assignment problem consists in finding the minimum cost perfect matching of a complete bipartite graph $G=(A \cup T,E)$, with $E=A \times T$ and $n=|A|=|T|$.
\begin{figure}[ht]
\begin{center}
\includegraphics[width=0.8\columnwidth]{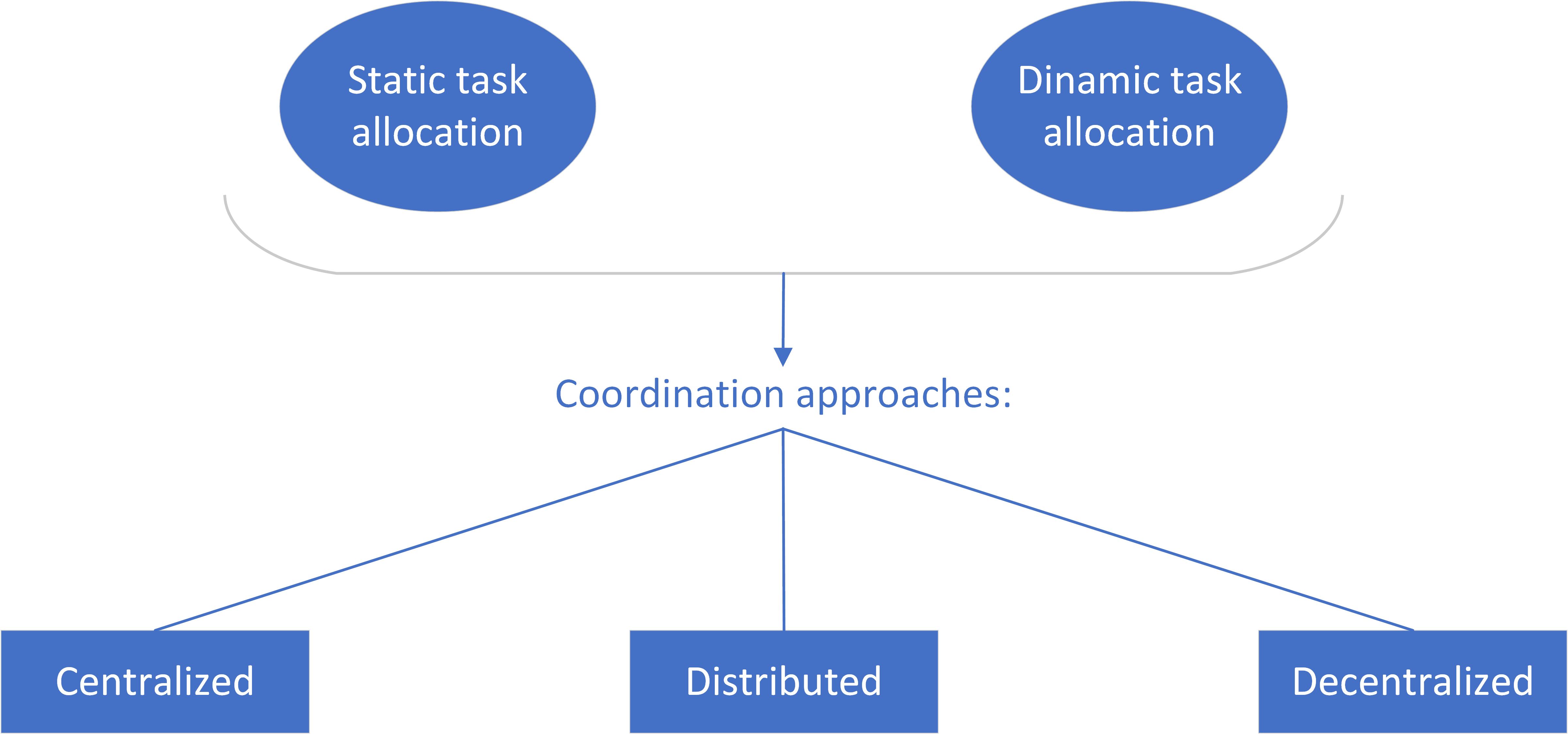}
\caption{Coordination approach framework for task allocation}
\label{SolutionApproaches}
\end{center}
\end{figure}

\subsection{Centralized coordination approaches}
There are a huge number of algorithms for the linear assignment problem (LAP). They can be subdivided into {\em primal}, {\em dual} and {\em primal-dual} algorithms. The worst-case time complexity of the best algorithms is $O(n^3)$.

We preliminary recall the mathematical formulation of the dual problem of the linear formulation of the LAP:

\begin{equation}\label{DAPobjective}
\max \sum_{i=1}^n u_i + \sum_{j=1}^n v_j
\end{equation}
subject to
\begin{equation}\label{dual_constraints}
u_i + v_j \leq c_{ij},\;\;\; \forall\;i,j \in \{1, \ldots, n\},
\end{equation}

\noindent where $u_i$ and $v_j$ are the (dual) variables.

\paragraph{Primal algorithms}
Primal algorithms are in general special implementations of the network simplex algorithm: one of the best primal algorithms is proposed in \cite{akgul1993genuinely} and runs in $O(n^3)$ time.

\paragraph{Dual algorithms}
Dual algorithms are iterative algorithms which at each iteration maintain a feasible dual solution and only at the final iteration they come up with a primal solution (i.e., a feasible assignment). In this regard also the primal-dual algorithms can be viewed as special dual algorithms. Typical dual algorithms are those based on successive shortest paths, signature, pseudoflow, interior point and auction methods. In the following, we concentrate on the auction methods because from the latter, one can easily derive distributed versions of the same.
For additional details, the reader is referred to \cite{burkard1999linear,burkard2009}.

For a short survey on the above solution algorithms for the LAP the reader is referred to a not so recent but detailed experimental comparison of some of the algorithms in \cite{dell2000algorithms}.
Another survey on the state of the art algorithms for the LAP is provided by \cite{burkard2009}.

\paragraph{Auction algorithms}
The first auction algorithm for the LAP was given by Bertsekas (1981) \cite{bertsekas1981new} and successively improved by Bertsekas and Eckstein \cite{bertsekas1988dual} through a scaling technique providing an algorithm that runs in $O(n^3 \log(nC))$, where $C = \max\{|c_{ij}|\}$. A survey of iterative combinatorial auction algorithms for task allocation in multi-agent systems can be found in, e.g., \cite{tang2010survey,bertsekas2009auction,jiang2015survey,scheffel2011experimental}.

The auction algorithm proposed by Bertsekas in \cite{bertsekas1981new} is an iterative algorithm that at each iteration maintains a triple $(x,(u,v))$ of primal and dual solutions that satisfy the complementary slackness conditions such that the dual solution is feasible. The algorithm terminates when also the corresponding primal solution is feasible. At each iteration,  the dual solution is updated and the corresponding primal solution (with respect to complementary slackness conditions) is found.

In particular, given a dual vector $v$, the optimal (feasible) dual vector $u$ can be obtained by considering
$u_i = \min_{j}\{c_{ij} - v_j\},$
\noindent and, hence, the dual problem can be rewritten as

\begin{equation}\label{DAPcompact}
\max q(v) = \sum_{i=1}^n \min_{j}\{c_{ij} - v_j\} + \sum_{j=1}^n v_j.
\end{equation}

\noindent Denoting with $j_i = \mbox{arg-min}_{j}\{c_{ij} - v_j\}$, the primal solution $x$, with $x_{i,j_i} = 1$ and 0 for $j \neq j_i$, with $i = 1, \ldots, n$, satisfies the complementary slackness conditions.

The dual problem has a nice economical interpretation. Assume that $p_j = -v_j$ represents the price that any agent will pay for being assigned to task $j$ and $u_i$ is the utility for agent $i$ for being assigned to a task. The dual assignment problem consists in determining $u_i$ and $p_j$ (i.e. $-v_j$) maximizing the agents' total net utility, such that agents' net utilities cannot be greater than the costs $c_{ij}$ they face. LP duality theory states that the maximum agents' total net utility equals the total assignment cost. At optimum, each task is assigned exactly to one agent, and the LP duality theory and complementary slackness conditions in particular assure that each agent $i$ is assigned to the most profitable task $j_i$, which guarantees that agent net utility $u_i - p_{j_i}$ is exactly equal to the assignment cost $c_{i,j_i}$.

From the LP duality theory applied to the AP,  we can derive the following auction algorithm \cite{bertsekas2009auction}. Assume that agents are assigned to tasks through a market mechanism, with agent $i$ acting according to its own best interest. Assume that task prices $p_j = -v_j$ are given. The total agent utility $(\sum_j u_j)$ is maximized if  we set each $u_j$ to its largest value allowed by the dual constraints, that is, $u_i = \min_{j}\{c_{ij} + p_j\}$. From the complementary slackness conditions, it follows that each agent $i$ will bid for the most profitable task $j_i$, i.e., with $c_{ij_i} + p_{j_i} = u_i$ in order to be assigned to it. If no task is bid by more than one agent, we reach an equilibrium and the assignment is optimal, otherwise we may change (increase) task prices $p_j$ in order to discourage agents to bid for the same task. This mechanism may be regarded as a naive auction algorithm that proceeds in rounds and halts if we get an equilibrium. We call it naive because it contains a flaw (as we will
show next), but it motivates a more sophisticated and correct algorithm.

At each round of the naive auction algorithm we start with a partial assignment and a given set of task prices and repeat the following two steps until all agents are assigned to their desired task (when we are at the equilibrium):
\begin{enumerate}
  \item {Bidding step}: Given task prices $p_j$ and a partial assignment of agents to tasks, (i) each unassigned agent $i$ bids for its most profitable task $j_i = \mbox{arg-min}_{j}\{c_{ij} + p_j\}$ with an offer equal to $p_{j_i} + \gamma_i$, with $\gamma_i = \beta_i - \alpha_i$, where $\alpha_i = \min_j \{c_{ij} + p_j\}$ and $\beta_i = \min_{j \neq j_i} \{c_{ij} + p_j\}$, while (ii) each already assigned agent still submits the previous winning bid (without changing their bid offers).
  \item {Pricing step}: Each task $j$ is assigned to the highest offering bidder (agent) for that target. The price $p_j$ of each task $j$ receiving a new (greater) bid is increased to the highest received offer, i.e., the new price value will be equal to $p_j + \gamma_i$.
\end{enumerate}

Unfortunately, this naive auction mechanism does not always work. It gets trapped in a cycle when (a) there is at least one unassigned agent and (b) each new winner bidder $i$ submitted an offer for its preferred task $j_i$ at its given target price $p_{j_i}$, i.e., $\gamma_i = 0$, meaning that its first and second best choices have the same cost.

In order to avoid this to happen, we need to keep rising the prices of tasks receiving new bids by at least a small amount $\epsilon > 0$. Therefore, we assume that agent $i$ will bid for its preferred task $j_i$ by offering $p_{j_i} + \gamma_i + \epsilon$.

This means that agent $i$ desires to be assigned to task $j_i$ if
$c_{ij_i} + p_{j_i} \leq \min_j \{c_{ij} + p_j\} + \epsilon = \alpha_i + \epsilon,$, which therefore is not necessarily its best choice. The above condition is known as $\epsilon$-complementary slackness (see, e.g., \cite{bertsekas2009auction}).

With this correction, the auction algorithm works ending in a finite number of rounds (depending on $\epsilon$), with each task receiving a bid. At the end, we are almost at an equilibrium with agent $i$ assigned to its almost desired task $j_i$. In general, this corresponds to an almost optimal solution for the assignment problem, since complementary conditions are only almost satisfied, while primal and dual complementary solutions are both feasible. It can be proved  that if the cost $c_{ij}$ are integers and $0 < \epsilon < \frac{1}{n}$, then the (corrected) auction algorithm ends with an optimal solution for the assignment problem (see, e.g., \cite{bertsekas2009auction}).

Without loss of generality, let us assume that $c_{ij} \geq 0$, and let $C = \max_{ij}\{c_{ij}\}$. In this case, it can be proved  that the auction algorithm runs in $O(n^3 \frac{C}{\epsilon})$ time (see, e.g., \cite{bertsekas2009auction}). Then, choosing $0 < \epsilon < \frac{1}{n}$, the algorithm returns an optimal solution in $O(n^4 C)$ time.
By using the scaling technique, Bertsekas and Eckstein in \cite{bertsekas1988dual} proposed a modified version of the above described auction algorithm that runs in $O(n^3 log(nC))$ time.
In real-world vehicle networks, the quality of solution in localized algorithms for task assignment is related with the communication network quality and range of communication. In \cite{lujak2011communication}, the influence of the communication range  and different strategies of movement on the task assignment value in the auction algorithm was evaluated in simulations in mobile (robot) agent task allocation scenarios.

\paragraph{Primal-dual algorithms}
Primal-dual algorithms start from a dual feasible solution $(u,v)$. From this solution, a restricted primal problem is defined and solved, consisting in finding the maximum cardinality matching on the bipartite subgraph $G^{\prime}=(A \cup T, E^{\prime})$, where $E^{\prime}=\{(i,j)\in E | c_{ij} - u_i - v_j = 0 \}$. If the optimal matching has size equal to $n$, we are done; otherwise, the dual solution is improved (the dual objective function is increased), while assuring that also the size of $E^{\prime}$ is increased, and the procedure is repeated.

Note that also the auction algorithms for LAP considers simultaneously primal and dual solutions but, differently from primal-dual algorithms, they can improve as well as worsen both the primal and the dual cost through the intermediate iterations, although at the end, the optimal assignment is found (see, e.g., \cite{bertsekas2009auction}).

\paragraph{Hungarian algorithm}

In particular the Hungarian algorithm, proposed by Munkres \cite{munkres1957algorithms} is a primal-dual algorithm. The original version of the algorithm runs in $O(n^4)$ time and was improved to $O(n^3)$ by Lawler in 1976 (see, e.g., \cite{lawler2001combinatorial}) by using successive shortest path technique when finding a new maximum cardinality matching after having updated the dual variables.

In the following we give some insights of the Hungarian algorithm that will be also useful for describing a decentralized version of the same. The Hungarian algorithm proceeds as follows:

\begin{itemize}
  \item Start with any feasible dual solution $(u,v)$, and any matching $M \subseteq E^{\prime}=\{(i,j)\in E | c_{ij} - u_i - v_j = 0 \}$. For the starting dual solution we can consider $v_j = \min_j\{c_{ij}\}$, with $j = 1, \ldots, n$, and $u_i = \min_i\{c_{ij} - v_j\}$, with $i = 1, \ldots, n$.
  \item While $M$ is not perfect repeat the following:
    \begin{enumerate}
    \item Given $M$ and $G^{\prime}=(A \cup T, E^{\prime})$, find an alternating augmenting path $P$ (i.e., a sequence of an odd number of edges that alternate edges of $E^{\prime}\backslash M$ and edges of $M$, starting and ending with non-matched edges); augment the matching by considering the new matching $M^{\prime} = M \backslash P \cup P \backslash M$. Note that $|M^{\prime}| = |M| + 1$.
        Update the matching $M$ (with $M^{\prime}$) and repeat until no new alternating augmenting path exists. $M$ is the maximum cardinality matching of $G^{\prime}$.
    \item If $M$ is not perfect, update the dual solution such that at least a new edge is added to the set of (admissible) edges $E^{\prime}=\{(i,j)\in E | c_{ij} - u_i - v_j = 0 \}$, and continue with a new iteration. In particular, we can achieve this result by updating the values of $u_i$ with $u_i + \delta$ and the values of $v_j$ with $v_j - \delta$, where $\delta = \min\{c_{ij} - u_i - v_j | i \in A^{\prime}, j \in T^{\prime}\}$ with $A^{\prime}$ and $T^{\prime}$ being the subsets of the vertices incident to the edges of the matching.
    \end{enumerate}
\end{itemize}

Searching for alternating augmenting path can be done by a graph visiting algorithm that identifies a forest of alternating trees of $G^{\prime}$.
Note that in each step of the loop we will either be increasing the size of $M$ or the size of $E^{\prime}$ so this process must terminate. Furthermore, when the process terminates, $M$ will be a perfect matching of $G^{\prime}=(A \cup T, E^{\prime})$, whose edge set $E^{\prime}$ is defined according to a feasible dual solution $(u,v)$. Since, the matching is perfect also for the complete bipartite graph $G$, the former represents a feasible primal solution for the assignment problem, respecting complementary constraints (by construction of of $E^{\prime}$), therefore the primal and dual solutions are optimal.

\paragraph{Parallel primal-dual algorithms}
A certain number of parallel algorithms for the linear assignment problem has been proposed. They are parallelized versions of primal-dual algorithms based on shortest path computations, of the auction algorithm, and of primal simplex-based methods. Among the most efficient parallel algorithms for the LAP is the one proposed by Orlin and Stein \cite{orlin1993parallel} that adopting cost scaling technique solves the problem using $\Omega(n^4)$ processors in $O(\log^3n \cdot \log(max\{cij\}))$ time.
For a review, the reader is referred to \cite{burkard2009,bertsekas2009auction,bertsekas1998network}.

\paragraph{Algorithms for the bottleneck assignment problem}
The bottleneck assignment problem can be solved in polynomial time for example by the so called {\em threshold algorithm} that alternates two phases (see, e.g., \cite{burkard2009,luss2012equitable}. In the first one, a threshold value $\bar{c}_{ij}$ is chosen and in the second phase, it is checked if the bipartite graph $G^{\prime}=(A \cup T,E^{\prime})$ admits a perfect matching or not, where $E^{\prime}=\{(i,j)\in E | c_{ij} \leq \bar{c}_{ij}\}$.

One possible way to implement the first phase is applying a binary search. This leads to a threshold algorithm that runs in $O(T(n) \log n)$ time, where $O(T(n))$ is the time complexity for perfect matching checking. One of the best time complexity algorithms by Punnen and Zhang  see, e.g., \cite{larusic2014asymmetric,punnen2009bottleneck} that runs in $O(m \sqrt{n \log n })$, where $m$ is the number of finite entries of the cost matrix $\{c_{ij}\}$.

\paragraph{Algorithms for the fair matching  problem}
The balanced assignment problem can be solved in polynomial time for example by means of an iterative algorithm based on a feasibility subroutine that runs in $O(k T(n)))$ (see, e.g., \cite{martello1984balanced}), where $k \leq n^2$ is the number of distinct values of $c_{ij}$ and $O(T(n))$ the time required to test if there is a feasible assignment on a subset $\bar{E} \subseteq E$ of the edges of the complete bipartite graph $G=(A \cup T,E)$. Testing if there is a feasible assignment on $\bar{E}$ corresponds to check if the bipartite graph $\bar{G}=(A \cup T,\bar{E})$ admits a perfect matching that can be done by solving the maximum cardinality matching of $\bar{G}$, e.g. in $O(n^{2.5})$ time \cite{hopcroft1973n}. Hence, since $k \leq n^2$, the overall algorithm run in $O(n^{4.5})$ time. Martello {\em et al.} in  \cite{martello1984balanced} improved the algorithm time complexity to $O(n^4)$ with a special refinement of the same.

\paragraph{Algorithms for an on-line bipartite matching}
Karp et al. in \cite{karp1990optimal} evaluate an on-line algorithm for bipartite matching by comparing its performance by the worst-case ratio of its profit to that of the optimal off-line algorithm.
They
propose an optimal online $1 - 1/e$ competitive simple randomized on-line
algorithm to maximize the size of the matching in an unweighted bipartite graph.
The best approximation algorithm for this problem is presented in \cite{menshadi2011offline} that applies the power of two choices paradigm, i.e., compute two offline matchings and use them to guide the adaptive online solutions.

Haeupler et al. in \cite{haeupler2011online} study the unrestricted weighted problem in the stochastic arrival model, and present the first approximation algorithms for it. They improve  $1-1/e$ -approximation for the online stochastic weighted matching problem  to a 0.667-approximation. Moreover, they apply a discounted LP technique to give an improved competitive algorithm for the online stochastic matching problem and  use the dual of the tightened LP to obtain a new upper bound on the optimal solution with a competitive ratio of 0.684.  Via   pseudo-matching, they obtain an algorithm with competitive ratio of 0.7036. They also present simple adaptive online algorithms to solve the online
(weighted) stochastic matching problem optimally for  the union of two matchings.

In \cite{manshadi2012online}, at each time step, a task is sampled independently from the given distribution and it needs to be matched upon its arrival to an agent. The goal is to maximize the number of allocations.
An online algorithm is presented for this problem with a competitive ratio of 0.702.  A key idea of the algorithm is to collect statistics about the decisions of the optimum offline solution using Monte Carlo sampling and use these statistics to guide the decisions of the online algorithm. The algorithm achieves a competitive ratio of 0.705 when the rates are integral.

\paragraph{In summary} While it is possible to solve most  of these problems using the simplex algorithm, each AP variation has specialized more efficient algorithms designed to take advantage of its special structure.

Many centralized algorithms have been developed for solving the assignment problem in  polynomial time (see, e.g., \cite{burkard2009}. One of the first such algorithms was the Hungarian algorithm \cite{munkres1957algorithms}. Other solution approaches include
augmenting path methods (see, e.g., \cite{jonker1987shortest,mills2007dynamic}), adaptations of the primal simplex method (see, e.g., \cite{orlin1997polynomial}), relaxation methods and auction algorithms (see, e.g., \cite{bertsekas2009auction}),
and signature methods (see, e.g., \cite{balinski1985signature}).

The complexity of the Hungarian method by using Fibonacci heaps   is $O(mn+n^2 \log n)$ \cite{fredman1987fibonacci}. Duan and Su's approach in \cite{duan2012scaling} give an algorithm whose running time  for integer weights   is $O(m\sqrt n \log N)$, where $m$ and $n$ are the number of edges and vertices and $N$ is the largest weight magnitude. Sankowski in \cite{sankowski2009maximum}
gave an $\tilde{O}(Wn^\omega)$\footnote[1]{$\tilde{O}$ denotes the so-called ``soft O'' notation} time, where $\omega$ is the matrix multiplication exponent, and $W$ is the highest edge
weight in the graph.

Duan and Pettie in \cite{duan2014linear} find an $O(m\epsilon^{-1}\log \epsilon^{-1}$ running time algorithm that computes  $(1 - \epsilon)$-approximate maximum weight matching for any fixed $\epsilon$.

Dell'Amico and Toth in \cite{dell2000algorithms} consider the classic linear assignment problem with a min-sum objective function, and the most efficient and easily available sequential codes for its solution that include: shortest path algorithms APC, CTCS, and LAPm; shortest augmenting path algorithm with reduction transfer procedure JV,  naive auction and sequential shortest path algorithm NAUCTION SP, two different implementations of the auction method, AFLP and AFR, and
pseudoflow cost scaling algorithm CSA.
Based on the results of the computational experiments  obtained on dense instances containing both randomly generated and benchmark problems,  it is not possible to obtain a precise ranking of the
eight algorithms. However,  APC is the fastest
code for the two cost class, and has a behavior, on average, similar to that of CTCS
for the other classes. Algorithm LAPm is the winner for the uniform random and
the geometric classes, and for the instances from the OR-library. No dominance with
respect to NAUCTION SP, CTCS and APC exists for the remaining classes. Code JV has a good and stable average performance for all the classes, and it is the best algorithm for
the uniform random (together with LAPm) and for the single-depot class. CSA performance strongly depends on the class, and it wins
for no-wait flow-shop classes.

\subsection{Distributed coordination approaches}
By distributed, we consider the algorithms that combine the concepts of centralized and decentralized coordination, and principally {\em market-based approaches}, where solutions are built based on a bidding-auctioning procedure between the bidders (agents) and   coordinators that play the role of auctioneers for allocating tasks to agents. There may be one or more coordinator agents as intermediaries in the task assignment process. The most known such algorithm is the auction algorithm that is presented in the following.

In this section we recall two distributed solution approaches respectively based on auction algorithm and on primal-dual Hungarian method.

The Bertsekas auction algorithm (see, e.g., \cite{bertsekas2009auction}) can be naturally implemented in a decentralized fashion. Zavlanos et al. \cite{zavlanos2008distributed} provide a distributed version of the auction algorithm proposed by Bertsekas for the considered networked systems with the lack of global information due to the limited communication capabilities of the agents. Updated prices, necessary for accurate bidding can be obtained in a multi-hop fashion only by local exchange of information between adjacent agents. No shared memory is available and the agents are required to store locally all the pricing information. This approach calculates the optimal solution in $O(\Delta n^3 C)$ time, with $\Delta \leq n - 1$ being the maximum network diameter of the communication network.

Another market-based algorithm has been proposed more recently by Liu and Shell in \cite{liu2013optimal}, that instead of auctioning via a series of selfish bids from customers (agents) adopts a mechanism from the perspective of a merchant. The algorithm is capable to produce a solution (equilibrium) that satisfies both merchant and customers and is globally optimal; its running time is $O(n^3 \log n)$.

Otte et al. in \cite{otte2019auctions} study various auction algorithms for task assignment in the multi-robot context, and study how lossy communication between the auctioneer and bidders affects solution quality. They demonstrate both analytically and experimentally that even though many auction algorithms have similar performance when communication is perfect, they degrade in different ways as communication quality decreases from perfect to nonexistent.  They compare six auction algorithms including: standard implementations of the Sequential Auction, Parallel Auction, Combinatorial Auction; a generalization of the Prim Allocation Auction called G-Prim; and two multi-round variants of a Repeated Parallel Auction. Variants of these auctions are also considered in which award information from previous rounds is rebroadcast by the auctioneer during later round. They conclude that the best performing auction changes based on the reliability of the communication between the bidders and the auctioneer.

Giordani {\em et al.} in \cite{giordani2010distributed,giordani2013distributed} propose a distributed version of the Hungarian method for solving the LAP, based on the concept of alternating augmenting paths, that are searched by maintaining a forest of alternating trees that is updated during the execution of the algorithm. In particular, given the current bipartite subgraph $G^{\prime}=(A \cup T, E^{\prime})$, where $E^{\prime}=\{(i,j)\in E | c_{ij} - u_i - v_j = 0 \}$, and $A$ and $T$ are agent and task vertices, respectively, the algorithm maintains forest $F_1$ of all the alternating trees rooted at free task vertices. Moreover, it maintains forest $F_2$ of the alternating trees of $G^{\prime}$ rooted at agent vertices containing all the agent/task vertices not contained in $F_1$. Clearly, the alternating trees in $F_2$ are not connected with vertices in $F_1$.

The algorithm involves root agents that initiate message exchange with other agents in the network via a depth-first search, and synchronize the decision rounds (iterations, each containing multiple communication hops) across all agents. Through autonomous calculations and the communication with the (agent) neighbors, with respect to the position of the vertex representing the agent in the spanning alternating forests, agents get and share the information about the position of each task vertex (whether in $F_1$ or $F_2$), the values of dual variables related to tasks, the value of $\delta$ for the dual variables' update, the new admissible edge entering in set of admissible edges of $G^{\prime}$ due to the dual variables' update, and the root agents $r(F_1)$ and $r(F_2)$ of forests $F_1$ and $F_2$ respectively. All these data are locally stored by each agent. In this way, there is no common coordinator or a shared memory of the agent's system. The agents, depending on the positions of the related vertices in the forests, change their roles, and accordingly execute some of the steps of the distributed Hungarian algorithm. The total computational time is $O(n^3)$ as well as the total number of messages exchanged by the robots; nonetheless, the computational time required to perform the local calculation by each robot is $O(n^2)$. Regarding the robustness of the proposed method, if the agent during the execution of the algorithm stops responding, it is considered erroneous and is eliminated from the further calculations. In the case where the agent was unmatched in forest $F_2$, the calculation continues without any modifications, ignoring the agent in question. Otherwise, the algorithm starts from the beginning excluding the same.

Chopra {\em et al.} in \cite{chopra2017distributed} propose a novel distributed version of the Hungarian method for solving the LAP that des not use any coordinator or shared memory. Specifically, each agent runs a local routine to execute ad-hoc substeps of the centralized Hungarian method and exchanges estimates of the solution with neighboring robots. The authors show that with their approach all agents converge to a common optimal assignment in a finite number ($O(n^3)$) of communication rounds if agents act synchronously. The overall performance of their approaches in terms of running time is only evaluated experimentally.

Eiselt and Marianov in \cite{eiselt2008employee} propose a model for the task assignment to employees with heterogeneous capabilities and multiple goals. Employees and tasks are mapped into the skill space where, after finding feasible matchings,  they are assigned to each other by minimizing employee-task distance to minimize assignment cost, boredom, and unfairness between employees' workloads.

Peters and Zelewski in \cite{peters2007assignment} develop two goal programming models for the employee assignment to workplaces according to both their competencies and preferences and the workplace requirements and attributes to ensure effective and efficient task performance.
A review and classification of the literature regarding workforce planning problems incorporating skills can be found in \cite{de2015workforce}.

The bottleneck assignment problem can be solved in polynomial time for example by the so called {\em threshold algorithm} that alternates two phases (see, e.g., \cite{burkard2009,luss2012equitable}. In the first one, a threshold value $\bar{c}_{ij}$ is chosen and in the second phase, it is checked if the bipartite graph $G^{\prime}=(A \cup T,E^{\prime})$ admits a perfect matching or not, where $E^{\prime}=\{(i,j)\in E | c_{ij} \leq \bar{c}_{ij}\}$.

One possible way to implement the first phase is applying a binary search. This leads to a threshold algorithm that run in $O(T(n) \log n)$ time, where $O(T(n))$ is the time complexity for perfect matching checking. One of the best time complexity algorithms by Punnen and Zhang  (see, e.g., \cite{larusic2014asymmetric,punnen2009bottleneck}) that runs in $O(m \sqrt{n \log n })$, where $m$ is the number of finite entries of the cost matrix $\{c_{ij}\}$.
Efrat et al. in \cite{efrat2001geometry} propose algorithms that, assuming planar objects, run in roughly $O(n^{1.5}\log n)$ time.
Pothen and Fan in \cite{pothen1990computing} propose a parallel algorithm with $O(nm)$ time complexity, which is currently among the best practical serial algorithms for maximum matching. However, its performance is sensitive to the order in which the vertices are processed for matching.

In \cite{azad2012multithreaded}, Azad et al. study the performance improvement of augmentation-based parallel matching algorithms for bipartite cardinality matching on multithreaded machines  over  serial algorithms and report extensive results and insights on efficient multithreaded implementations of three classes of algorithms based on their manner of searching for augmenting paths: breadth-first-search, depth-first-search, and a combination of both.

In \cite{efrat2001geometry}, algorithms for the balanced assignment problem and minimum deviation assignment are presented that run in roughly $O(n^{10/3}$ and, as such, are more efficient than
the algorithms of \cite{martello1984balanced} and \cite{gupta1988minimum} that run in $O(n^4)$ time  on general bipartite graphs.
Kennington and Wang in \cite{kennington1992shortest} present  a shortest augmenting path algorithm for solving the semi-assignment problem  in which  each iteration during the final phase of the procedure (also known as the end-game) obtains an additional assignment.

\subsection{Decentralized coordination approaches}
In contrast to centralized and distributed coordination approaches to task allocation where full knowledge of global information is assumed available to every relevant decision maker (central decision maker  or fleet coordinator (fleet owner) and (vehicle) bidder agents), in the decentralized task assignment approaches, there is no coordinator and each vehicle agent disposes only of its local (possibly incomplete and imperfect) information and finds its local assignment  based exclusively on this information and the communication with the rest of the agents and interaction with its environment.

In general, decentralized approaches
have several advantages, i.e., real-time property, robustness, and scalability. These characteristics are in general absent in centralized and distributed approaches that  outperform decentralized approaches in terms of efficiency especially for large-scale instances.
The decentralized decision-making   does not include any intermediary. In case of imperfect  communication, conflicts may occur. This is why
the related literature in decentralized multi-vehicle cooperative control is related with consensus, i.e., the agreement of all vehicles on some common features by negotiating with their local neighbors. General consensus issues are related with, e.g., positions, velocities, and attitudes.
In the following, we analyze  localized,  scalable, and decentralized heuristic algorithms for coordination of  deterministic and dynamic task assignment in open vehicle fleets. We concentrate on the approaches resulting    both in task assignment  feasibility and efficiency even though  these approaches usually have no quality of solution guarantees.

Decentralized task assignment approaches have been mostly developed in the multi-robot and Unmanned Aerial Vehicle (UAV) coordination domain. The most known ones are  sequential auction-based or consensus and negotiation-based algorithms (e.g., \cite{nunes2015multi,choi2009consensus,johnson2011asynchronous}).

One of the most known approaches for the decentralized task assignment in the coordination of a fleet of unmanned vehicles when all-to-all inter-vehicle communication  is not possible  is the Consensus-Based Auction Algorithm (CBAA) and its more general version that allows for the assignment of bundles of tasks to each agent called the Consensus-Based Bundle Algorithm (CBBA) \cite{choi2009consensus}.

The CBAA is a polynomial time market-based decentralized task selection agreement protocol running in two phases: in the first phase,  each vehicle places a bid on a task asynchronously with the rest of the fleet, and in the second, consensus phase, conflicting assignments are identified and resolved through local communication between neighboring agents within certain predefined rules to avoid task conflicts.
The agents use a consensus strategy to converge on the list of winning bids and use that list to determine the winner and associated winning scores. The list  accounts for inconsistent information among agents guaranteeing a conflict-free assignment for all.
This allows conflict resolution over all tasks that is robust  to inconsistencies in the situational awareness across the fleet and  the changes in the communication network topology. If the resulting scoring scheme satisfies a diminishing marginal gain  property (i.e., the value of a task does not increase as other tasks are assigned to the same agent before it), a feasible, conflict-free solution is guaranteed.

Provided that the scoring function abides by the principle of diminishing marginal gains, the CBBA has convergence guarantees. In a synchronized conflict resolution phase over a static communication network, it produces the same solution as the sequential greedy algorithm  sharing across the fleet the corresponding winning bid values and winning agent information.  Moreover, the convergence time is bounded from above and it does not depend on the inconsistency in the situational awareness over the agent set.

In \cite{choi2009consensus}, it is analytically shown that CBAA produces the same solution as some centralized sequential greedy procedures, and this solution guarantees 50$\%$ optimality.
Segui-Gasco et al. \cite{segui2015decentralised} propose a decentralized algorithm for multi-robot task allocation  with
a constant factor approximation of 63 $\%$
 for positive-valued monotone submodular utility functions, and of 37 $\%$ for   general positive-valued submodular utility functions.
Therefore, the authors improve the approximation guarantee of Choi et
al. \cite{choi2009consensus} for monotone positive-valued submodular utility functions from 50$\%$ to 37$\%$.

The CBBA has also been extended to consider coupled constraints \cite{choi2010decentralized,whitten2011decentralized}.
Choi et al. in \cite{choi2010decentralized} extended CBBA for heterogeneous task allocation to UAV agents with different qualifications and various cooperation constraints. The CBBA was extended with task decomposition and a scoring modification to allow for soft-constraints related with cooperation preferences and a decentralized task elimination protocol that ensures satisfaction of the hard-constraints related with cooperation requirements. The performance of the algorithms was analyzed in Monte-Carlo simulations in some randomly generated experiments.

The CBBA was also extended in \cite{whitten2011decentralized} to consider
the assignment of tasks with assignment constraints and also with different types of coupled and temporal constraints, where it was assumed   that assigned tasks are  executed in the order defined by their temporal precedence.

The Temporal Sequential Single-Item auction (TeSSI) algorithm \cite{nunes2015multi} allocates tasks with time windows to cooperative robot agents using a variant
of the sequential single-item auction algorithm.
Contrary to the CBBA algorithm that does not let the change of the
start time of the tasks once they are allocated, and thus reduces
the number of tasks that the algorithm allocates, the TeSSI algorithm overcomes this limitation by allowing tasks’ start times
to change, which results in higher allocation rates.

The main features of the TeSSI algorithm are a fast and systematic processing of temporal constraints and two bidding methods that optimize either completion time or a combination of completion time
and distance.
The main objective function used in the TeSSI algorithm is the makespan (the time the last task is finished)even though it is also combined with  total distance traveled.
Each robot maintains temporal consistency of its
allocated tasks using a simple temporal network.   The authors show that TeSSI outperforms a baseline greedy algorithm and the  CBBA through random experiments and related work datasets.

Ponda et al. in \cite{ponda2010decentralized} further extend the CBBA  to tasks with time windows and address re-planning in dynamic environments and consider agents with different capabilities. Agents obtain new plans  based on the changes in the environment considering new tasks while pruning older or irrelevant ones.

One of the drawbacks of the CBBA algorithm is that it relies on global synchronization mechanisms which  are hard to enforce in decentralized environments.
Johnson et al.  \cite{johnson2011asynchronous} proposed the asynchronous CBBA (ACBBA) for agents that communicate asynchronously. To allow for asynchrony in communication, the  ACBBA contains a set of local deconfliction rules that do not require access to the global information.
In ACBBA, agents locally replan their actions that, possibly, affect only a limited number of agents.

Johnson et al. \cite{johnson2013hybrid} propose a situational awareness algorithm for task assignment when agents   predict the bids of their neighbors, in order to obtain more informed decisions in a cooperative way.

To respond to the problem with local information consistency assumption that reduces optimization capabilities compared to  global information assumption approaches, Johnson et al.
\cite{johnson2017decentralized}  proposed a  Bid Warped Consensus-Based Bundle Algorithm that converges for all deterministic objective functions and has nontrivial performance guarantees for submodular and some non-submodular objective functions. They  analyse convergence and performance of the algorithm and show its efficiency compared with some other relevant local and global information approaches.

Another extension to the CBBA is provided by Binetti et al. \cite{binetti2013decentralized} that consider the decentralized surveillance problem by a team of robots. Tasks are assigned to each robot with the additional constraint that a subset of the tasks called critical tasks must be assigned.  The authors use the CBBA incorporating hard constraints in order to ensure that the critical tasks are not left unassigned.

In \cite{garcia2015cooperative}, Garcia and Casbeer present a robust task assignment algorithm that reduces communication between vehicles in uncertain environments.  Piece-wise optimal decentralized allocation of tasks is considered for a group of unmanned aerial vehicles. They present a framework for multi-agent cooperative decision making under communication constraints.  Each vehicle estimates the position of all other vehicles in order to assign tasks based on these estimates, and it also implements event-based broadcasting strategies that allow the multi-agent system representing the vehicle fleet to use communication resources more efficiently. The agents implement a simple decentralized auction scheme in order to resolve possible conflicts.

Cui et al. in \cite{cui2013game}  investigate game  theory-based  negotiation for task  allocation  in the multi-robot task assignment context.   Tasks are initially allocated using a Contract Net (see \cite{smith1980contract}) -based approach, after which,   a   negotiation approach  employing the utility functions to  select  the  negotiation  robot agents and construct the negotiation set is proposed.  Then, a  game  theory-based  negotiation  strategy   achieves the Pareto-optimal solution for the task reallocation. Extensive simulation results  demonstrate the efficiency of such a task assignment approach.

Yet another extension of the Consensus-Based Bundle Algorithm (CBBA) allowing for the fast allocation of new tasks without a full
reallocation of existing tasks is CBBA with Partial Replanning (CBBA-PR) \cite{buckman2019partial}. The algorithm enables the multi-agent system to trade-off between convergence time and increased coordination by resetting a portion of their
previous allocation at every round of bidding on tasks. By resetting the last tasks allocated
by each agent, the convergence of the MAS to a conflict-free solution is assured.
CBBA-PR can be further improved by reducing the team size involved in the replanning, further
reducing the communication burden of the team and runtime of CBBA-PR.

In \cite{sayyaadi2010distributed}, Sayyaadi and Moarref investigate a proportional task assignment problem in which it is desired for (robot) agents to have equal duty to capability ratios, i.e., the agents with more capability should perform more tasks. They address this problem as a combination of deployment and consensus problems in which agents should reach consensus over the value of their duty to capability ratios. They propose a distributed, asynchronous and scalable algorithm for this problem in continuous time domain.

Duran et al. in \cite{duran2017} study the problem of finding the list of solutions with strictly increasing cost for the Semi-Assignment Problem. Four different algorithms are described and compared. The results show that they find the exact list of solutions, and considerably reduce the computation times in comparison with the other exact approaches.

Spivey et al. in \cite{spivey2015} propose a distributed, flexible, and scalable control scheme that evenly allocates tasks.  Dynamic load balancing exploits feedback information about the status of tasks and vehicles with the objective to keep  a balanced task load and, thus, force cooperation in the solution of the randomized bottleneck task assignment problem.

In summary, most of the state-of-the-art decentralized and deterministic coordination approaches for task allocation are  heuristic algorithms developed for    multi-robot or UAV task allocation  scenarios that often include both operational and tactical constraints of a vehicle fleet and its environment. Even though their adaptation for the use in open vehicle fleets does not seem difficult, it remains an open challenge, especially if we consider  task allocation efficiency, the key performance indicator of commercial open fleets.

\section{Challenges in open vehicle fleet coordination}\label{Challenges}
In this paper, we proposed new mathematical programming models of dynamic
versions of the following assignment problems well-known in combinatorial optimization and applicable in open vehicle fleets:
the assignment problem, bottleneck assignment problem, fair matching problem, dynamic minimum deviation assignment problem,  $\sum_{k}$- assignment problem, the semi-assignment problem, the assignment problem with side constraints,  and the assignment problem while recognizing agent   qualification.
The goal of the studied  problems   is finding an optimal (minimum cost or maximum profit) assignment to the (vehicle) agents of the   tasks that are known at the time of decision-making. These approaches do not take into account unknown tasks that may appear once when the current tasks are completed.

With the long term objective of decentralizing and democratizing shared mobility, we categorized solution approaches for static and dynamic task assignment problems applicable in open vehicle fleets  into centralized, distributed, and decentralized and discussed their main characteristics.
The presented distributed and decentralized task assignment methods are   applicable in distributed and decentralized open vehicle fleets, respectively. In case of decentralized fleets, the issues related with privacy, trust, and control intrinsic to centralized systems are gone.

We focused on homogeneous vehicle agents and tasks, i.e., each vehicle agent is able to complete each task with equal efficiency but varying cost or profit. In the real world that might not be the case since in open vehicle fleets, the vehicles tend to be heterogeneous. The proposed mathematical programs can easily be adapted to this case by varying the agent-task assignment cost/profit depending on the performance efficiency of an agent; in case of an agent inapt to perform a task, its agent-task assignment cost is assigned a very large value.

With fully decentralized scalable coordination of task allocation, there is no need to put limits to the size of the system. However, even though scalable task allocation and related coordination mechanisms are essential for efficiently managing large-scale open vehicle fleet systems, it should be noticed that, for real-world applications, they  need to be complemented with scalable and efficient solution approaches to other combinatorial optimization problems depending on the context, e.g., dial-a-ride problem, traveling salesperson problem, etc.

We dealt with the deterministic and dynamic assignment problem where real-time reassignment is beneficial since both the variables and parameters of the optimization problem are perfectly known at each period. However, when dealing with real-world stochastic environments with increased sensor noise, a too high frequency of task re-assignment may result in a churning effect in the assignment and may lead to increased human errors. Thus, a chosen coordination method must consider churning in this context to obtain good overall task allocation performance (see, e.g., \cite{alighanbari2008robust}).

A truly open vehicle fleet system should work also based on heterogeneous software agents produced by multiple producers. The agent software could be an open source and/or there may be multiple proprietary software companies working on a common open fleet coordination standard.
The Agreement Technologies (AT) paradigm \cite{791} identifies and relates various such technologies. It provides a sandbox of mechanisms to support coordination among (heterogeneous) autonomous software agents, which focuses on the concept of agreement between them. To this respect, AT-based systems not only support the interactions for reaching agreement in a coordinated manner (e.g. as part of a distributed or decentralized algorithm) but are also endowed with means to specify and govern the ``space'' of agreements that can be reached, as well as monitoring agreement execution. In particular, in truly open vehicle fleet systems where there may be a multitude of (possibly heterogeneous) software providers, semantic mismatches among vehicle agents need to be dealt with through the alignment of ontologies, so that vehicle agents can reach a common understanding on the elements of agreements.

Furthermore, (weak) constraints on agreement and agreement processes (often also called \emph{norms}) need to be defined and represented in a declarative manner, so autonomous agents can decide as to whether they will adopt them, determine as to how far they are applicable in a certain situation, dynamically generate priorities among conflicting norms depending on the context, etc. In addition, trust and reputation models are necessary for keeping track of whether the agreements reached, and their executions, respect the requirements put forward by norms and organisational constraints. So, norms and trust can be conceived as a priori and a posteriori approaches, respectively, to support the security in relation to the coordination process. How to find seamless and effective means of integrating the different distributed and decentralized algorithms outlined in this paper in such a framework is still an open issue that we will treat in our future work.

The presented distributed and decentralized coordination methods  for dynamic task assignment may be applied to semi-autonomous and autonomous vehicles and are  a necessary part of reaching full vehicle fleet autonomy. They may not fix the mobility concerns, but they will definitely improve them as they are directly related to giving a higher control both to an individual driver (or to an autonomous vehicle) and to a customer (rider).  Intrinsically, these methods aid in changing the hierarchical tree structure of the transportation networks to a more horizontal one. Indirect benefits of such coordination methods, among others, include  higher efficiency, smaller carbon footprints and less traffic jams.
In the long run, they will facilitate   more decentralized, autonomous, and transparent open vehicle fleets, but above all, they will further  the task allocation efficiency and fair rewards and benefits of vehicles, drivers, customers, and riders  proportional to  their participation  in large and open fleets.

\paragraph{Acknowledgements} This work has been partially supported  by E-Logistics project financed by  the French Agency for Environment and Energy Management (ADEME) and   ``IntelliEDGE'' project  ``RTI2018-095390-B-C33'' funded by Spanish Ministry of Science, Innovation and Universities.
\bibliographystyle{elsarticle-num}
\bibliography{Survey.bib}
\endgroup
\end{document}